# Interference effects on the $I_D/I_G$ ratio of the Raman spectra of diamond-like carbon thin films

Elitza Petrova, Savcho Tinchev, and Petranka Nikolova,

Institute of Electronics, Bulgarian Academy of Sciences,
72 Tzarigradsko Chaussee, 1784 Sofia, Bulgaria

Abstract:

Ratio of the intensity of D- Raman peak and G- Raman peak ($I_D/I_G$) is often used for characterization of diamond-like carbon films, for example to estimate number and size of the $sp^2$ clusters. In this paper we investigate how the interference effects in the film influence this ratio. It is shown that for films on silicon substrates the distortion of the real ratio can reach 10 % and for films on metals even 40 %. This should be taken into account if some conclusions are made from the measured $I_D/I_G$ ratio.

1. Introduction.

Raman spectroscopy is a widely used method to characterize carbon-based materials [1-3]. Basic structural properties, such as nanocluster size can be derived from the Raman measurements. In amorphous diamond-like carbon thin films Raman spectroscopy is probably the most important method for diagnostics of the material [4-7]. Raman spectra of these films usually show two distinct peaks – the so called D- and G- peaks. The parameters of the two peaks (position, width and intensity ratio) are used for the characterization of the thin film material. For example, an increase in $I_D/I_G$ ratio is ascribed to an increase in the number and/or the size of $sp^2$ clusters.

However, if the film thickness is comparable to the wavelength of the laser /Raman light, interference effects in the film take place and the measured Raman spectra can be disturbed. In the present paper the interference effect on the intensity ratio of the D- and G- peaks of the Raman signal is calculated. It is shown that in diamond-like carbon (DLC) films with thickness of some hundred nanometers this effect should be taken into account if some conclusions are made from the $I_D/I_G$ ratio.

Nemanich et al. [8] were the first who used the interference enhancement of the Raman signal from very thin titanium and titanium oxide films. Their system consisted of a thin sample layer on the top of a dielectric spacer layer, which in turn was deposited on a highly reflecting metallic film. For an optical thickness of one quarter of the laser wavelength, a destructive interference for the reflected laser beam occurs while constructive interference occurs for the Raman signal. Thus interference enhancement of the Raman signal is possible. The concept of interference enhanced Raman scattering was used later by Ramsteiner et al. [9, 10] to find exact theoretical expressions including absorption, multiple reflections, and interference effects for the intensity of Raman scattering from thin films. They verified their calculations by experimentally measured Raman spectra of thin amorphous hydrogenated carbon films deposited on crystalline Si substrates. In contrast to the configuration from Nemanich et al., an a-C:H film on a Si surface acts both as a sample layer and a dielectric spacer layer. As a result, instead of the expected monotonic increase of the Raman intensity up to film thicknesses equal to the laser penetration depth, interference effect was observed. Strong oscillatory behavior of the Raman intensity of the main G-peak was predicted and experimentally verified.

However, the optical thickness of the investigated layer is different for each Raman frequency. Therefore the $I_D/I_G$ ratio will be different for films with different thicknesses even if these films have the same structure. We expect this effect to be not very strong, especially for a-C:H films on Si substrates. Nevertheless, it should be taken into account if some conclusions are made from the measured $I_D/I_G$ ratio.

2. Results and discussions.

We investigate theoretically hydrogenated amorphous carbon thin films deposited on two different substrates – silicon and metal (Al). Our calculations are made in a manner similar to the calculation in [10]. For the amplitude of the laser beam in a certain depth *x* taking into account its multiple reflections within the film and interference at the interface air/film we have:

$$E_L = const \frac{t_1 e^{-i\beta_1 x} + t_1 r_2 e^{-i\beta_1 (2d-x)}}{1 + r_1 r_2 e^{-i2d\beta_1}} e^{i\omega_L t} \qquad (1)$$

Here $\beta_1 = (2\pi/\lambda_1)(n_1 - ik_1)$ is the wavenumber for the laser light, $(n_1 - ik_1)$ is the complex refractive index of the DLC film taken at the laser wavelength and *d* is the thickness of the sample layer. Calculated for the amplitude of the wave, $t_1$ is the transmission coefficient at the interface air/film and $r_1$ and $r_2$ are the reflection coefficients at the interface air/film and interface film /substrate, respectively.

Similarly, the Raman scattering amplitude could be expressed as:

$$E_R = const \frac{t_2 e^{-i\beta_2 x} + t_2 r_2 e^{-i\beta_2 (2d-x)}}{1 + r_1 r_2 e^{-i2d\beta_2}} e^{i\omega_R t} \qquad (2)$$

Here $\beta_2 = (2\pi/\lambda_2)(n_2 - ik_2)$ is the wavenumber of the Raman light, $(n_2 - ik_2)$ is the complex refractive index of the DLC film, $t_2$ is the amplitude transmission coefficient at the interface between the film and the air and $r_1$ and $r_2$ are the respective amplitude reflection coefficients. Each of these parameters is taken at a particular Raman wavelength.

The total Raman signal was calculated by:

$$I = const \frac{\mathrm{Re}\{n_2^L\}}{\mathrm{Re}\{n_2^R\}} \int_0^d |E_L|^2 |E_R|^2 \, dx \,, \qquad (3)$$

where $\mathrm{Re}\{n_2^L\}$ and $\mathrm{Re}\{n_2^R\}$ are the real parts of the sample layer's refractive indices at the laser and Raman wavelengths, respectively.

The exact theoretical expression of the Raman signal as a function of the layer thicknesses and the optical constants found in [10] is complicated. Therefore the Eq. (3) was evaluated numerically. The refractive indices used in calculations are given in Tables 1 – 4.

Fig. 1 shows Raman signal intensities $I_G(d)$ of the G-peak and $I_D(d)$ of the D-peak calculated as a function of the sample thickness $d$ for a DLC film with a typical refractive index [11] on a Si substrate. The calculations are carried out for three different laser wavelengths 514.5 nm, 633 nm and 458 nm. The positions of Raman peaks [12] are slightly different in dependence of the laser excitation wavelength. As expected the interference effects are more pronounced for 633 nm, because the attenuation of the Raman components is much stronger for shorter wavelengths. From the curves in Fig. 1 it is difficult to see the difference between G-peak and D-peak intensities corresponding to a certain laser wavelength because the wavelengths of G- and D-peaks are close to each other. However, the differences are more pronounced in the plotted relations of $I_D/I_G$ ratio on the film thickness $d$ shown in Figs. 2 - 4.

Obviously, depending on the film thickness the real $I_D/I_G$ ratio can be distorted up to about 10% for DLC films on Si substrates. On metals, however, for example on Al (see Fig. 5), which is the case in the solar thermal absorbers, large oscillations could be expected and distortion up to 40 % could significantly influence the $I_D/I_G$ ratio and therefore the conclusion made from it.

Substrate effects and Raman spectra of hydrogenated amorphous carbon films deposited onto silicon and other substrates were investigated in [13]. In general, intensity increases with thickness for all substrates. Significant increase in intensity was observed on metallic substrates. However, an increasing film thickness beyond about 100 nm leads to a reduction in intensity. Such behavior can be seen in our calculated Raman signal intensity on Al substrate – Fig. 6. Similarly, the addition of the thermally evaporated Al layer on a Si substrate [14] increases the measured Raman intensity by a factor of 2 for 514.5 nm excitation, 40 for 325 nm excitation and 10 for 244 nm excitation. This enhancement was explained by SERS (surface enhanced Raman spectroscopy) as the thermally evaporated Al is not a flat surface, but an assembly of blob-like and rod-like structures of 50–150-nm size (as seen by SEM). To our opinion all these observations can be explained by interference enhanced Raman scattering as our calculations show.

## 3. Conclusions

In this paper it is shown that the $I_D/I_G$ ratio often used to characterize DLC films can be influenced by interference effects in the DLC film. The interference effects in the DLC films are more pronounced for longer laser wavelengths where the distortion of the $I_D/I_G$ ratio can reach 10 % for DLC films on silicon substrates. For films on metal substrates the distortion of this ratio can be even much more and can heavily compromise conclusions about the film microstructure.

<u>Figures captions:</u>

Fig. 1. Calculated Raman G-peak (a) and D-peak (b) as a function of the sample thickness *d* for a DLC film on a Si substrate. The calculations are carried out for three laser lengths 514.5 nm, 633 nm and 458 nm.

Fig. 2. Calculated distortion of $I_D/I_G$ ratio caused by interference for DLC film on a Si substrate and laser wavelength of 458 nm.

Fig. 3. Calculated distortion of $I_D/I_G$ ratio caused by interference for DLC film on a Si substrate and laser wavelength of 514.5 nm.

Fig. 4. Calculated distortion of $I_D/I_G$ ratio caused by interference for DLC film on a Si substrate and laser wavelength of 633 nm.

Fig. 5. Calculated distortion of $I_D/I_G$ ratio caused by interference for DLC film on aluminum substrate and laser wavelength of 633 nm.

Fig. 6. Calculated Raman intensity of the G-peak and D-peak as a function of the sample thickness *d* for a DLC film on aluminum substrate and laser wavelength of 633 nm.

Table 1: The refractive indices used in calculation for laser wavelength of 633 nm on Si substrate

| | $n_{DLC}$ | $k_{DLC}$ | $n_{Si}$ | $k_{Si}$ |
|---|---|---|---|---|
| $\lambda_G$ = 698.9 nm | 2.176 | 0.034 | 3.774 | 0.011 |
| $\lambda_D$ = 686.2 nm | 2.183 | 0.038 | 3.791 | 0.012 |
| $\lambda$ = 633 nm | 2.209 | 0.07 | 3.879 | 0.016 |

Table 2: The refractive indices used in calculation for laser wavelength of 514.5 nm on Si substrate

| | $n_{DLC}$ | $k_{DLC}$ | $n_{Si}$ | $k_{Si}$ |
|---|---|---|---|---|
| $\lambda_G$ = 558.6 nm | 2.236 | 0.127 | 4.044 | 0.026 |
| $\lambda_D$ = 550 nm | 2.241 | 0.134 | 4.077 | 0.028 |
| $\lambda$ = 514.5 nm | 2.25 | 0.17 | 4.216 | 0.037 |

Table 3: The refractive indices used in calculation for laser wavelength of 458 nm on Si substrate

| | $n_{DLC}$ | $k_{DLC}$ | $n_{Si}$ | $k_{Si}$ |
|---|---|---|---|---|
| $\lambda_G$ = 492.7 nm | 2.254 | 0.195 | 4.331 | 0.048 |
| $\lambda_D$ = 486.6 nm | 2.254 | 0.202 | 4.369 | 0.052 |
| $\lambda$ = 458 nm | 2.254 | 0.238 | 4.577 | 0.077 |

Table 4: The refractive indices used in calculation for laser wavelength of 633 nm on Al substrate

| | $n_{DLC}$ | $k_{DLC}$ | $n_{Al}$ | $k_{Al}$ |
|---|---|---|---|---|
| $\lambda_G$ = 698.9 nm | 2.176 | 0.034 | 1.821 | 8.3 |
| $\lambda_D$ = 686.2 nm | 2.183 | 0.038 | 1.72 | 8.182 |
| $\lambda$ = 633 nm | 2.209 | 0.07 | 1.373 | 7.618 |

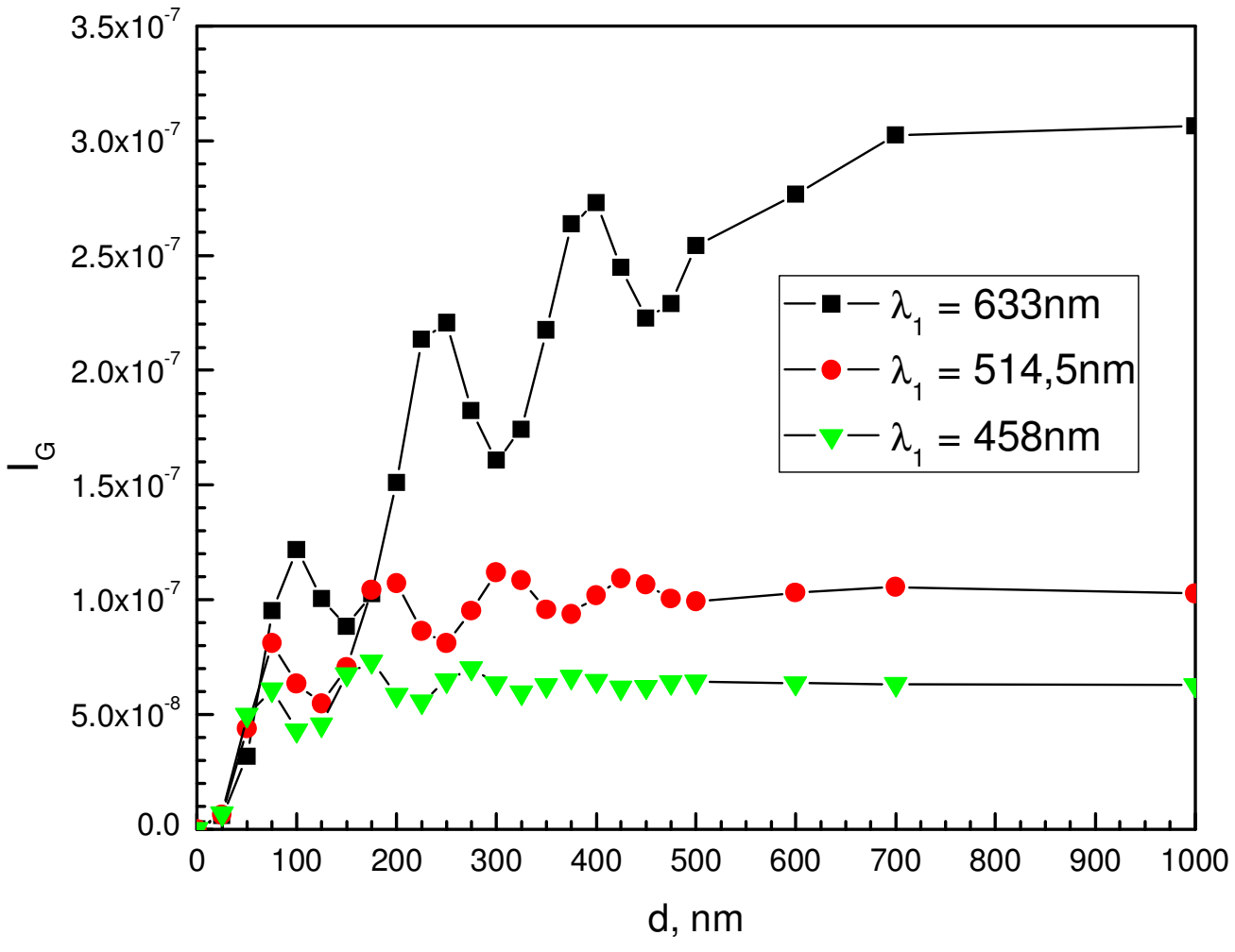

Fig, 1a

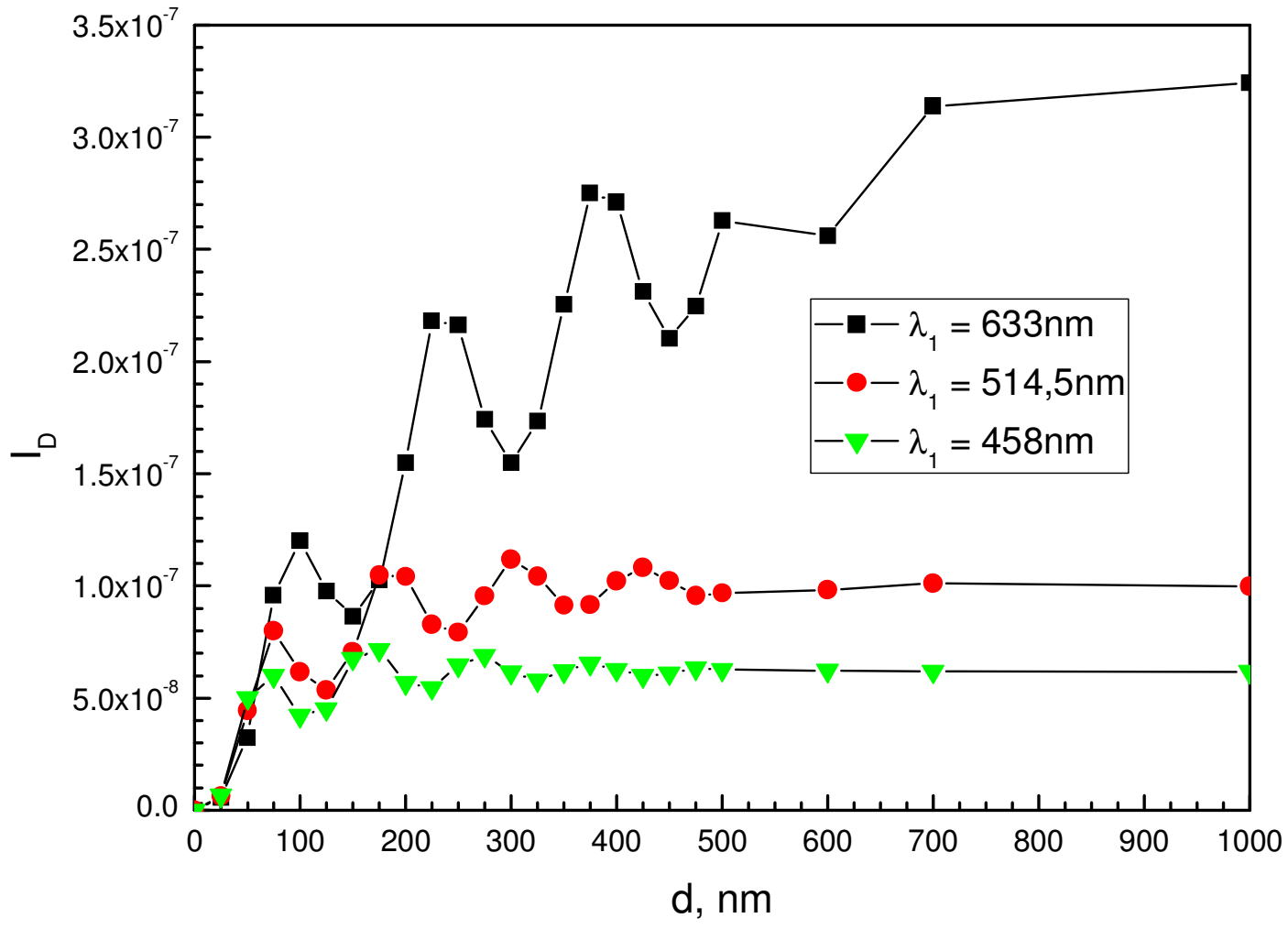

Fig. 1b

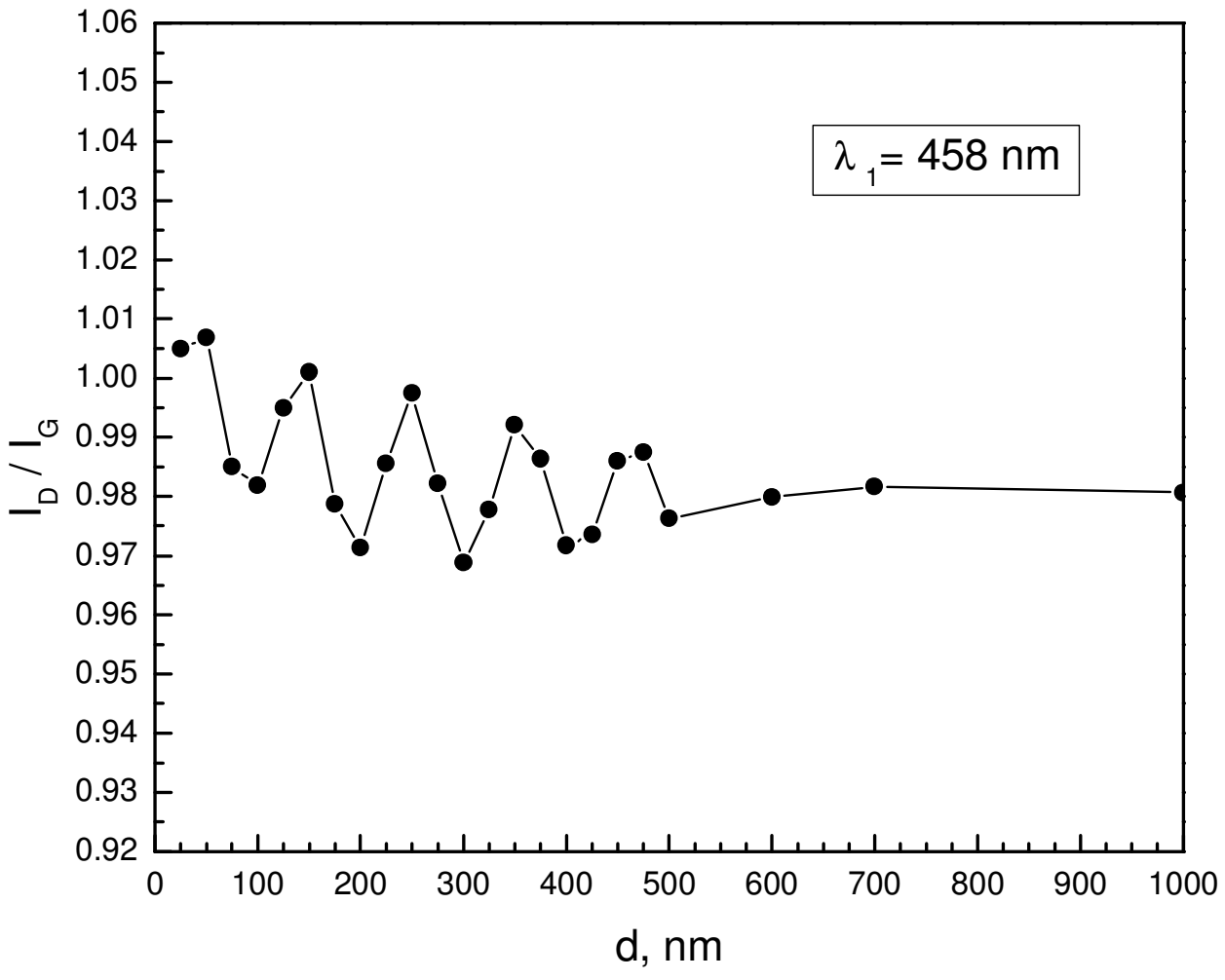

Fig. 2

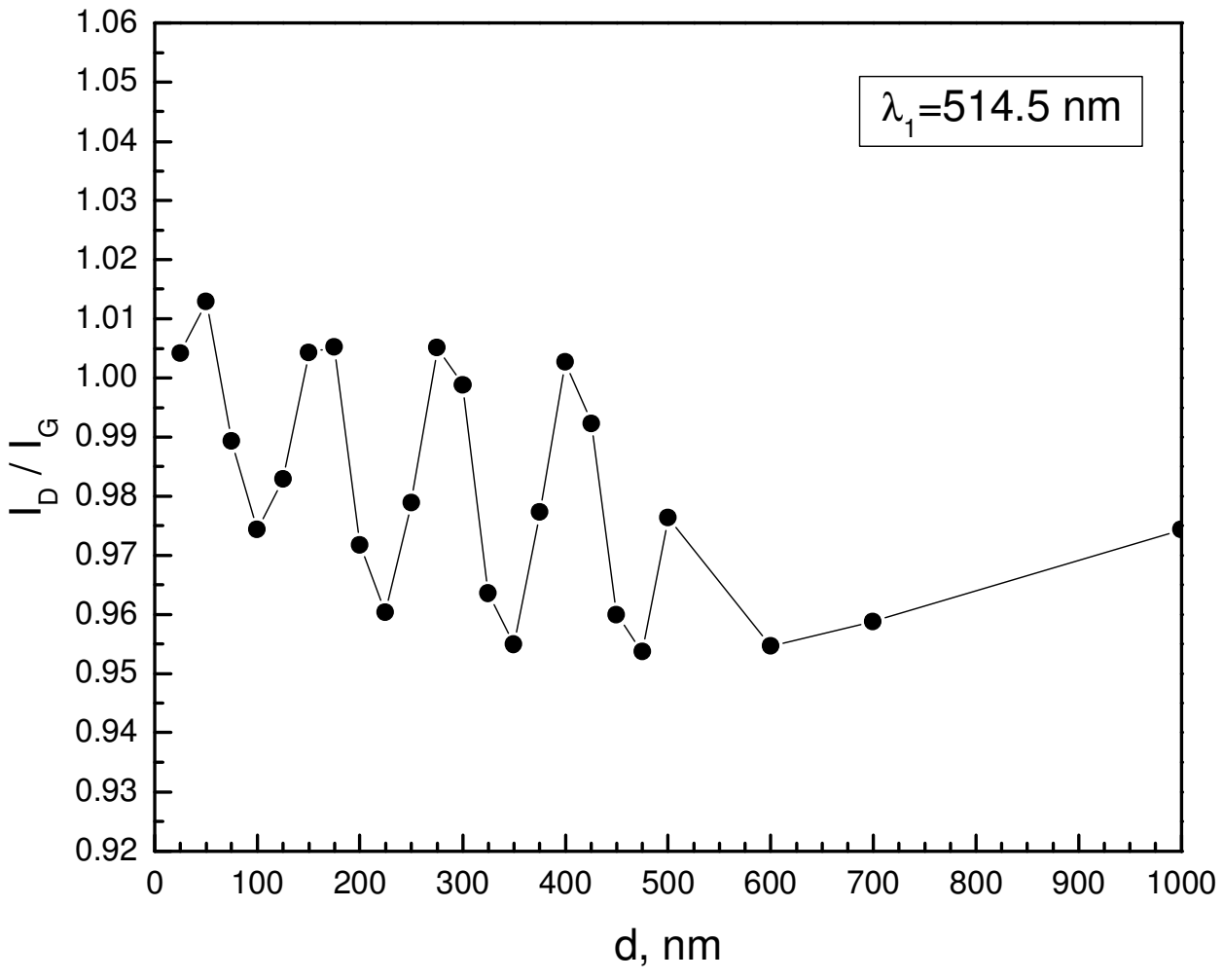

Fig. 3

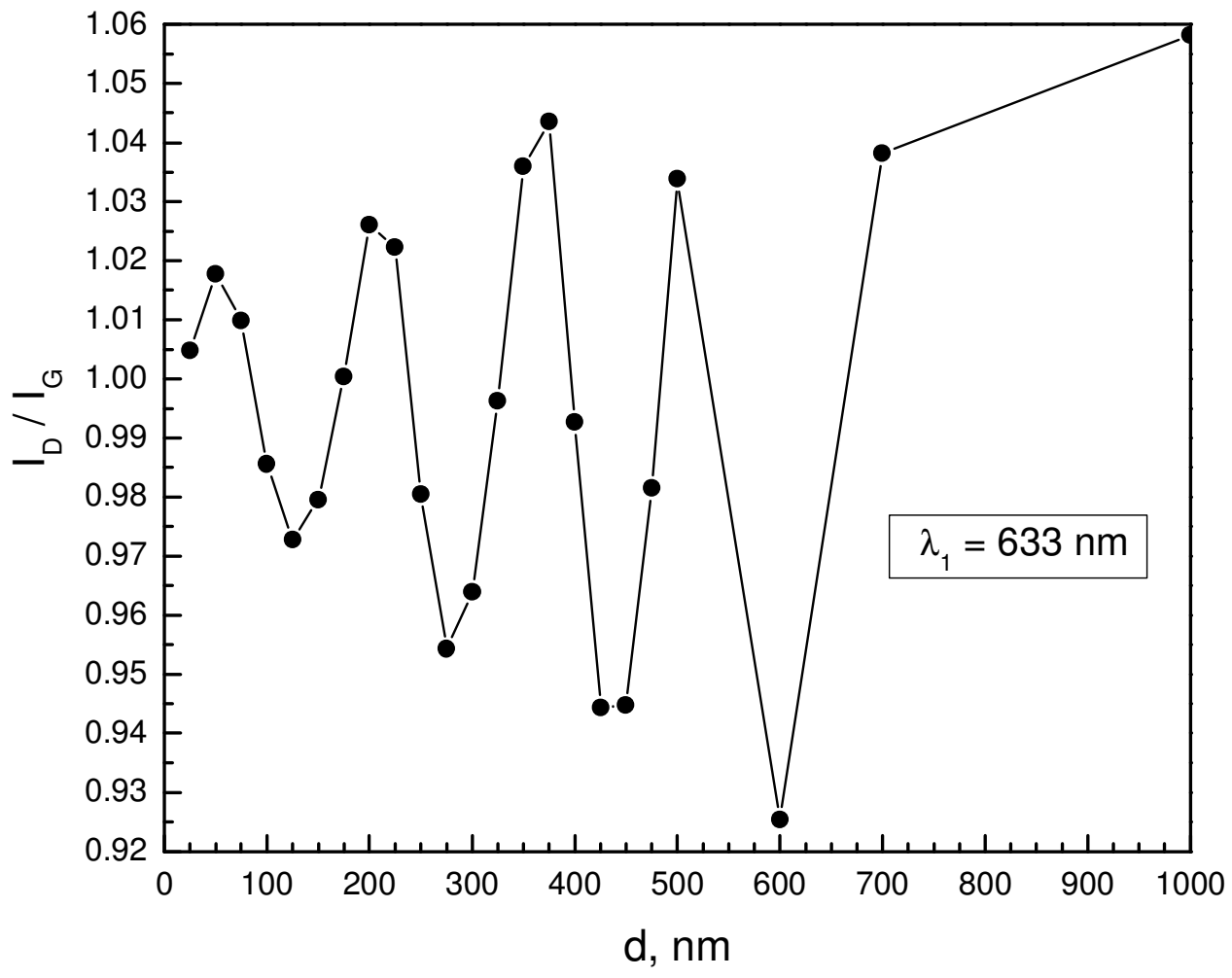

Fig. 4

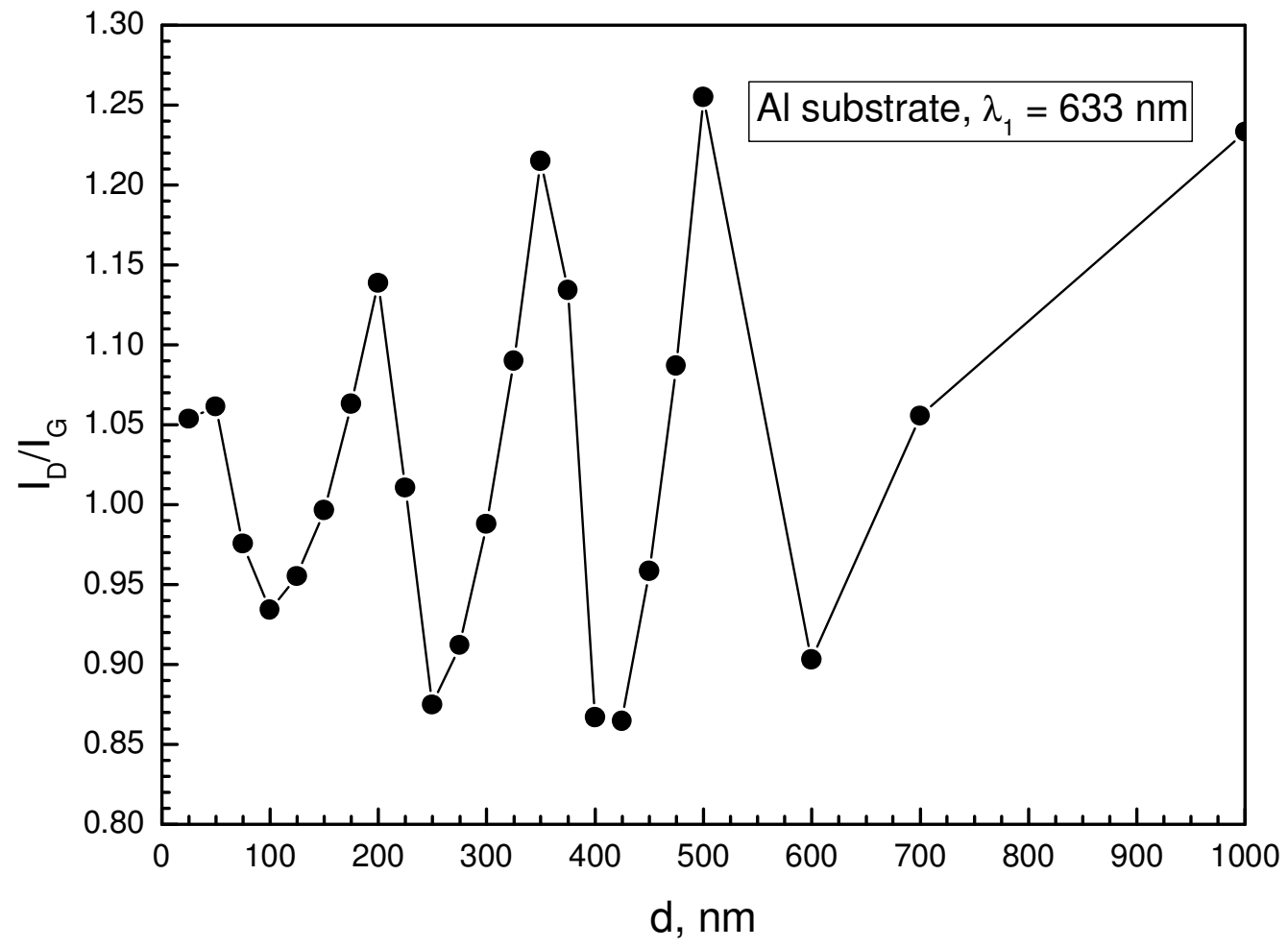

Fig. 5

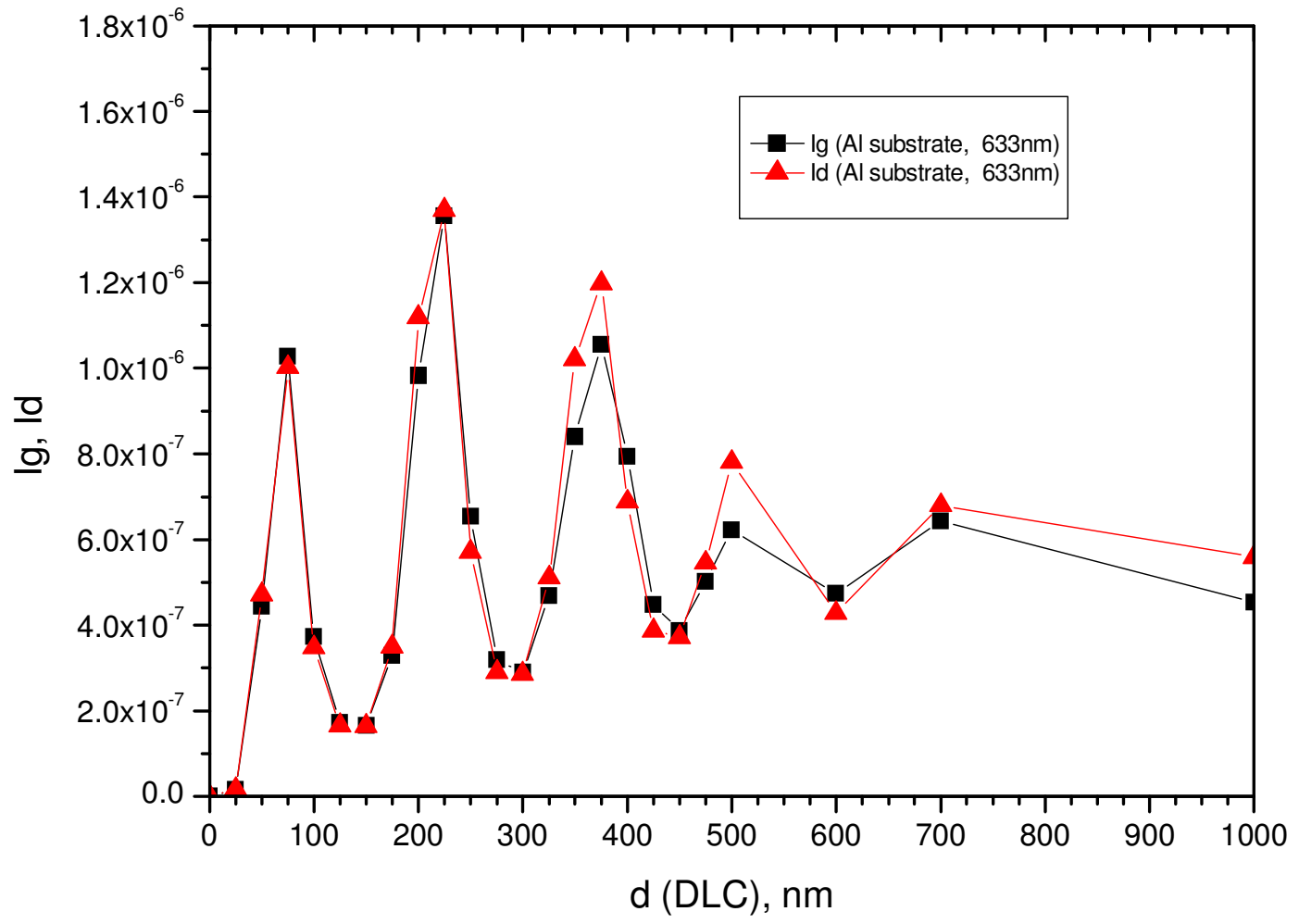

Fig. 6